\documentclass[a4paper,fleqn,usenatbib]{mnras}

\usepackage{graphicx}
\usepackage{amsmath,amssymb,amsfonts}

\newcommand{\be}{\begin{equation}}
\newcommand{\ee}{\end{equation}}
\newcommand{\bel}[1]{\begin{equation}\label{#1}}
\newcommand{\ba}{\begin{eqnarray}}
\newcommand{\ea}{\end{eqnarray}}
\newcommand{\bal}[1]{\begin{eqnarray}\label{#1}}

\def\apj{Astrophysical Journal}                
\def\apjl{Astrophysical Journal}             

\def\apjs{ApJS}              
\def\mnras{MNRAS}            
\def\aap{A\&A}                
\def\pasa{Publications of the Astron.~Soc.~of Australia}

\usepackage[usenames]{color}

\begin{document}

\title{Estimates of black-hole natal kick velocities from observations of low-mass X-ray binaries}
\author[Ilya Mandel]
{Ilya Mandel$^{1}$\thanks{E-mail: imandel@star.sr.bham.ac.uk}\\
$^{1}$School of Physics and Astronomy, University of Birmingham, Birmingham, B15 2TT, United Kingdom}

\pagerange{\pageref{firstpage}--\pageref{lastpage}}\pubyear{2015}

\maketitle

\label{firstpage}


\begin{abstract}

The birth kicks of black holes, arising from asymmetric mass ejection or neutrino emission during core-collapse supernovae, are of great interest for both observationally constraining supernova models and population-synthesis studies of binary evolution.  Recently, several efforts were undertaken to estimate black hole birth kicks from observations of black-hole low-mass X-ray binaries.  We follow up on this work, specifically focussing on the highest estimated black-hole kick velocities.   We find that existing observations do not require black hole birth kicks in excess of approximately 100 km/s, although higher kicks are not ruled out.
\end{abstract}

\begin{keywords}
stars: black holes, binaries: close, X-rays: binaries, Galaxy: kinematics and dynamics
\end{keywords}

\section*{Introduction}

Unlike neutron-star supernova-induced birth kicks, which are constrained from pulsar proper motion observations \citep{Hobbs:2005}, black-hole birth kicks are poorly known \citep[e.g.,][]{Repetto:2012}.  They are, nonetheless, of great interest to stellar and binary astrophysicists.  Black-hole birth kicks play a key role in the evolution of massive stellar binaries and predictions of compact-binary merger rates observable through their gravitational-wave signatures \citep[e.g.,][]{ratesdoc}.  For example, in a population-synthesis study of binary evolution, \citet{Dominik:2012} find that varying black-hole kicks from zero to the distribution used for neutron stars decreases the rate of binary black hole mergers by two orders of magnitude, as a significant fraction of potential binary black hole systems are disrupted by large supernovae kicks.  Birth kicks are also of interest in supernova modeling as constraints on the supernova explosion mechanism.  For example \citet{Janka:2013} uses the high kicks inferred by \cite{Repetto:2012} to support a supernova model in which a neutron star is initially formed and converted into a black hole after a delay during which ejecta fall back asymmetrically, contributing further momentum to the black hole.

Black-hole X-ray binaries provide a promising tool for measuring black-hole birth kicks.  Typically, measuring the birth kicks requires the knowledge of the position and velocity of the X-ray binary in the Galaxy.  This information makes it possible to integrate the trajectory of the binary backwards in the Galactic potential until it intersects the Galactic plane, where the binary is assumed to have been born.  This assumption is consistent with both the observed birth locations of massive stars in a thin disk in the Galactic plane \citep[e.g.,][]{Urquhart:2014}, and more specifically the finding that metal abundances in the secondary in black-hole X-ray binaries is consistent with a thin-disk origin \citep[e.g.,][]{GonzalezHernandez:2008}.  Population synthesis modeling can then be applied to include the contribution of Blaauw kicks \citep{Blaauw:1961} from mass loss during the supernova and constrain the range of supernova-induced natal kicks required to preserve the binary with current orbital parameters.  

This approach has been applied to several black-hole X-ray binaries.  \citet{Willems:2005} find that a birth kick of a few tens of km/s (and no more than around 200 km/s) was given in the supernova that formed GRO J1655-40.  \citet{Wong:2012} find that Cygnus X-1 likely received a non-zero birth kick (cf.~\citep{Nelemans:1999}), but with a velocity no larger than around $80$ km/s.  \citet{Wong:2014} find that the birth kick of IC10 X-1 was smaller than $130$ km/s.  \citet{Fragos:2009} find the largest black-hole birth kick inferred from systems with both position and velocity measurements in XTE J1118 + 480, which they conclude must be between $80$ and $310$ km/s.

However, the number of black-hole X-ray binaries with accurate position and velocity measurements is very limited \citep{MillerJones:2014}.  This motivated \citet{Repetto:2012} and \citet{RepettoNelemans:2015} to extend the approach above to the analysis of black-hole low-mass X-ray binaries with a position measurement but without a 3-dimensional velocity measurement.  They use population-synthesis models of binary evolution to argue that, without natal supernova kicks, such binaries cannot achieve overall initial velocities in excess of $\sim 50$ km/s -- the maximum velocity that could be attributed to the Blaauw kick from mass loss during a supernova, as larger mass ejection would disrupt the binary.   

\citet{RepettoNelemans:2015} estimate the minimum initial velocity of the binaries from their current observed position in Galactic spherical coordinates $(\rho,z)$ as follows.  They consider a binary that starts out on the Galactic plane and moves directly out of the Galactic plane, with an initial velocity pointed perpendicular to the Galactic plane, until it reaches the present observed location.  They use the difference in the Galactic potential at the observed spherical coordinate position $(\rho,z)$ and assumed starting position $(\rho,z=0)$ to infer the minimal initial velocity of the binary from the conservation of energy  (see Sec.~3.3 of \citet{RepettoNelemans:2015}):
\begin{equation}
v_\textrm{initial}=\sqrt{2 [\Phi(\rho,z)-\Phi(\rho,0)]}.
\end{equation}

\citet{RepettoNelemans:2015} analyze seven black-hole low-mass X-ray binaries.  They find that several of them must have had initial velocities in excess of what is possible without natal kicks, most notably H 1705-250 (for which they estimate an initial binary velocity between $360$ and $440$ km/s) and, to a lesser extent, XTE J1118+480 (for which they estimate an initial binary velocity of $\gtrsim 70$ km/s), with lower initial velocities for other systems.   This provides evidence for significant natal kicks.  

We follow up on this work below, particularly focusing on the ultra-fast kick required for H 1705-250.  We re-evaluate the required initial velocities, and consider the contribution of uncertainties in the observations and modeling, in order to critically evaluate the claimed evidence for black hole kicks of order $\sim 400$ km/s.  We conclude that existing observations do not require the existence of black-hole birth kicks above $\sim 100$ km/s.

\section*{Initial velocity estimation}

Eq.~(1) converts the extra energy necessary to get the binary up to the observed height above the Galactic plane into a constraint on the initial kinetic energy, and, hence, the natal velocity of the binary following the BH supernova.   For example, Eq.~(1) evaluated using the Galactic potential described by model 2 of \citet{Irrgang:2013} (see below) indicates that the potential difference between the claimed location of H 1705-250, ($\rho=0.5$ kpc, $z=1.35$ kpc), and its projection onto the Galactic plane ($\rho=0.5$ kpc, $z=0$ kpc), corresponds to a requirement of $v_\textrm{initial} \geq 370$ km/s on the initial velocity.   

However, the potential difference between the observed location and its projection onto the Galactic plane is not an accurate indication of the necessary initial velocity.  The gradient of the potential has a significant component in the direction perpendicular to the Galactic plane, so a displacement in this direction is energetically unfavorable and requires a high initial velocity.  Moreover, a purely vertical displacement does not correspond to a physical solution: a velocity that is initially perpendicular to the Galactic plane would not, in general, yield a purely vertical displacement in the Galactic potential.  In fact, if the potential difference between initial and observed location were sufficient to determine the necessary initial velocity, it would have been potential to start the binary at a location with a potential similar to that at its present location; for example, starting the binary ($\rho=3.5$ kpc, $z=0$ kpc) would yield a required natal velocity of only $10$ km/s.  

On the other hand, motion along equipotential lines is not, in general, possible.  An additional complication is that the Galactic disk, in which the binary is assumed to be born (see below), is rotating with a large velocity roughly independent of radius, $\sim 225$ km/s; therefore, we must consider the initial velocity of the binary relative to the velocity of the disk.  In effect, the need for conservation of angular momentum in the $\hat{z}$ direction means that in order to find a binary such as H 1705-250 close to a position directly above the Galactic center, at $\rho \to 0$, it must first receive a kick roughly canceling its azimuthal velocity if it was born in the disk at larger cylindrical radii, thereby setting the $\hat{z}$ component of its angular momentum to zero.  This kick should be accompanied with a smaller component necessary to move to high Galactic latitudes, so we expect that a minimal initial velocity just above $\sim 225$ km/s would be required.
 
This is what we find when we determine the minimum initial velocity by integrating the binary's motion backwards in the Galactic potential from its current observed position, noting the binary's velocity at crossings of the Galactic plane, $z=0$, under the assumption that the binary was born in the plane of the disk.   Performing this integration requires specifying the unknown velocities at the observed position of the binary.    A possible trajectory passing through ($\rho=0.5$ kpc, $z=1.35$ kpc) is illustrated in Figure 1.  The velocity at the Galactic plane crossing is then compared to the Keplerian orbital velocity of the disk in which the binary is born.   Minimizing the initial velocity relative to the disk for this position via a grid over the 3-dimensional velocity at the observed location yields an initial velocity of only $v_\textrm{initial}  \approx 230$ km/s, only $60\%$ of the required $v_\textrm{initial} \geq 370$ km/s estimated from Eq.~(1).  The binary navigates this trajectory in well under 1 Gyr,  consistent with the possible age of an LMXB.

\begin{figure}
\includegraphics[width=\columnwidth]{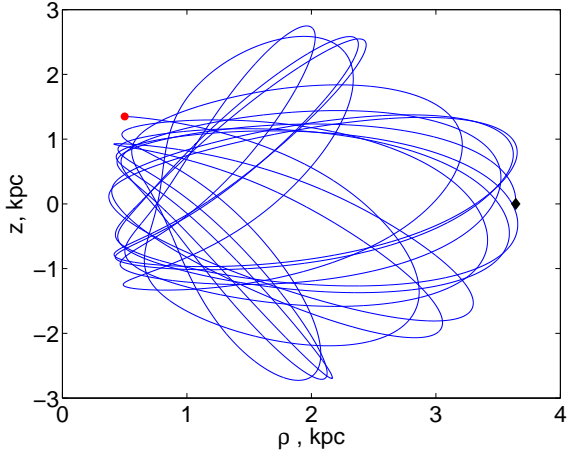}
\caption{\label{fig:trajectory}
The trajectory of a binary in the Galactic potential, integrated backwards from the claimed observed position of H 1705-250, ($\rho=0.5$ kpc, $z=1.35$ kpc), indicated with a red circle.   Motion in the azimuthal $\phi$ direction is not shown.  At the Galactic plane crossing indicated by the black diamond, the total velocity relative to the Keplerian disk is $230$ km/s.}
\end{figure}

\section*{Uncertainty in the observed position}

The distance to low-mass X-ray binaries is difficult to measure precisely.  \citet{Remillard:1996} provide distance estimates of $6$---$10$ kpc for H 1705-250, which are used by \citet{RepettoNelemans:2015} to provide error bars on the initial binary velocity (these error bars do not correctly account for the full range of uncertainty \citep{Repetto:2015}). Meanwhile, \citet{Martin:1995} claim distance estimates of $2$---$8.4$ kpc.  If the lower boundary of $2$ kpc is taken from the \citet{Martin:1995} estimate, the binary H 1705-250 could be located only $\sim 0.3$ kpc away from the Galactic plane.  This is consistent with the scale height of the thick disk naturally arising from dynamical relaxation processes, obviating the need for a natal kick.

\section*{Initial location of the binary}
The scale height of the Galactic disk is $\sim 0.3$ kpc.  Since the Galactic potential has a minimum in the Galactic plane among all points at a fixed cylindrical coordinate $\rho$, it requires the most kinetic energy for the binary to rise from an initial location in the Galactic plane.  Therefore, assuming that the binary is initially located away from the Galactic plane reduces the initial velocity required to reach an observed height away from the Galactic plane.  

However, massive stars which can give rise to a black hole in a low-mass X-ray binary appear to be born in a much thinner disk of only $20$ to $30$ pc \citep[e.g.,][]{Urquhart:2014}.  Variations in initial position by only a few tens of parsecs do not lead to an appreciable change in the required initial velocity.  On the other hand, a somewhat larger scale height -- and, critically, a lower azimuthal velocity -- may be possible in the bulge itself; if H 1705-250 is really located above the bulge, this could reduce the necessary orbital velocity by a further $\sim 10$ -- $20\%$, depending on assumptions.\footnote{Under the most optimistic assumptions, H 1705-250 may have been born near its present location in $(\rho, z)$, making a black-hole birth kick unnecessary; however this appears relatively unlikely given the present-day massive-star formation.}


\section*{Uncertainty in the Galactic potential.}  

The Galactic mass density and potential \citep{Paczynski:1990} are not known perfectly.  Most of the uncertainty is confined to the halo, despite improved measurements which rely on masers and hypervelocity stars \citep[e.g.,][]{Irrgang:2013}.  Therefore, we do not expect the uncertainty in the Galactic potential to significant impact our findings.  We use the three models presented in \citep{Irrgang:2013}, all of which fit the observed data well, to estimate that uncertainty in the potential models should contribute at most at the $\sim 10$ -- $20\%$ level to the inferred binary velocities.

\section*{Conclusion}

We have discussed several sources of uncertainty in the inference on black-hole birth velocities from the observed positions of black-hole low-mass X-ray binaries.  Other complications include the possibility that the initial kick comes from a dynamical interaction (and subsequent ejection of the binary from its host cluster) rather than a supernova-induced kick \citep{Repetto:2012}.  We also note that a small fraction of binaries could get significant kicks of many tens to a few hundred km/s from strong 3-body interactions in the Galactic disk, with interlopers approaching to within the semimajor axis of the binary.  However, we estimate that this mechanism would apply to $\lesssim 0.1\%$ of all binaries, and so is an unlikely explanation for the observed systems.

When accurate position and velocity measurements of low-mass X-ray binaries are available \citep{MillerJones:2014}, the binary's trajectory can be integrated backward in the Galactic potential to deduce the velocity of the binary at birth, assuming that it was born in the Galactic plane.  This calculation for XTE J1118+480, which yields a minimal birth kick of 80 km/s \citep{Fragos:2009}, provides a firm lower bound on the maximum black hole kick.  

In the absence of such a measurement, statistical inference is, in principle, possible.  For example, the small (few tens of km/s according to \citep{Remillard:1996}) line-of-sight velocity of H 1705-250 might suggest that a high total current velocity is less likely.  The small number of observations and the uncertain selection effects make reliance on statistical arguments precarious. 
A statistical analysis of all observed black-hole X-ray binaries with position measurements was undertaken by \citep{Repetto:2012}, who considered a sample of 16 black-hole X-ray binaries to argue that their distribution was consistent with black holes receiving similar birth kicks to neutron stars.   However, this conclusion again largely relied on two systems requiring particularly strong kicks, most notably H 1705-250.    

We found that the potential difference between the current observed position of a low-mass X-ray binary and the projection of that position onto the Galactic plane is not an accurate conservative estimate of the minimal initial velocity of the binary.  Moreover, inference on initial velocities is significantly compromised by the uncertainty in observed positions of some LMXBs.  We conclude that existing 
observations of only the spatial locations of low-mass X-ray binaries are not sufficient to confidently deduce the existence of strong black-hole supernova kicks of order $400$ km/s, as claimed by \citep{RepettoNelemans:2015}.   Upcoming high-precision position and velocity observations, such as those enabled by Gaia, should improve our ability to infer black-hole birth kicks.

\vspace{0.1in} 

We thank Serena Repetto, Chris Belczynski, Melvyn Davies, Cole Miller, Gijs Nelemans, Steinn Sigurdsson, Ian Stevens, and Simon Stevenson for comments and discussion.  We especially thank Selma de Mink for a critical reading of the manuscript.

\vspace{-0.4in}
\bibliographystyle{mnras}
\bibliography{Mandel}

\begin{thebibliography}{}
\makeatletter
\relax
\def\mn@urlcharsother{\let\do\@makeother \do\$\do\&\do\#\do\^\do\_\do\%\do\~}
\def\mn@doi{\begingroup\mn@urlcharsother \@ifnextchar [ {\mn@doi@}
  {\mn@doi@[]}}
\def\mn@doi@[#1]#2{\def\@tempa{#1}\ifx\@tempa\@empty \href
  {http://dx.doi.org/#2} {doi:#2}\else \href {http://dx.doi.org/#2} {#1}\fi
  \endgroup}
\def\mn@eprint#1#2{\mn@eprint@#1:#2::\@nil}
\def\mn@eprint@arXiv#1{\href {http://arxiv.org/abs/#1} {{\tt arXiv:#1}}}
\def\mn@eprint@dblp#1{\href {http://dblp.uni-trier.de/rec/bibtex/#1.xml}
  {dblp:#1}}
\def\mn@eprint@#1:#2:#3:#4\@nil{\def\@tempa {#1}\def\@tempb {#2}\def\@tempc
  {#3}\ifx \@tempc \@empty \let \@tempc \@tempb \let \@tempb \@tempa \fi \ifx
  \@tempb \@empty \def\@tempb {arXiv}\fi \@ifundefined
  {mn@eprint@\@tempb}{\@tempb:\@tempc}{\expandafter \expandafter \csname
  mn@eprint@\@tempb\endcsname \expandafter{\@tempc}}}

\bibitem[\protect\citeauthoryear{Abadie et~al.}{Abadie et~al.}{2010}]{ratesdoc}
Abadie J.,  et~al., 2010, \mn@doi [Classical and Quantum Gravity]
  {10.1088/0264-9381/27/17/173001}, \href
  {http://adsabs.harvard.edu/abs/2010CQGra..27q3001A} {27, 173001}

\bibitem[\protect\citeauthoryear{{Blaauw}}{{Blaauw}}{1961}]{Blaauw:1961}
{Blaauw} A.,  1961, Bull.~Astron.~Inst.~Netherlands, \href
  {http://adsabs.harvard.edu/abs/1961BAN....15..265B} {15, 265}

\bibitem[\protect\citeauthoryear{{Dominik}, {Belczynski}, {Fryer}, {Holz},
  {Berti}, {Bulik}, {Mandel}  \& {O'Shaughnessy}}{{Dominik}
  et~al.}{2012}]{Dominik:2012}
{Dominik} M.,  {Belczynski} K.,  {Fryer} C.,  {Holz} D.~E.,  {Berti} E.,
  {Bulik} T.,  {Mandel} I.,   {O'Shaughnessy} R.,  2012, \mn@doi [\apj]
  {10.1088/0004-637X/759/1/52}, \href
  {http://adsabs.harvard.edu/abs/2012ApJ...759...52D} {759, 52}

\bibitem[\protect\citeauthoryear{{Fragos}, {Willems}, {Kalogera}, {Ivanova},
  {Rockefeller}, {Fryer}  \& {Young}}{{Fragos} et~al.}{2009}]{Fragos:2009}
{Fragos} T.,  {Willems} B.,  {Kalogera} V.,  {Ivanova} N.,  {Rockefeller} G.,
  {Fryer} C.~L.,   {Young} P.~A.,  2009, \mn@doi [\apj]
  {10.1088/0004-637X/697/2/1057}, \href
  {http://adsabs.harvard.edu/abs/2009ApJ...697.1057F} {697, 1057}

\bibitem[\protect\citeauthoryear{{Gonz{\'a}lez Hern{\'a}ndez}, {Rebolo},
  {Israelian}, {Filippenko}, {Chornock}, {Tominaga}, {Umeda}  \&
  {Nomoto}}{{Gonz{\'a}lez Hern{\'a}ndez} et~al.}{2008}]{GonzalezHernandez:2008}
{Gonz{\'a}lez Hern{\'a}ndez} J.~I.,  {Rebolo} R.,  {Israelian} G.,
  {Filippenko} A.~V.,  {Chornock} R.,  {Tominaga} N.,  {Umeda} H.,   {Nomoto}
  K.,  2008, \mn@doi [\apj] {10.1086/586888}, \href
  {http://adsabs.harvard.edu/abs/2008ApJ...679..732G} {679, 732}

\bibitem[\protect\citeauthoryear{{Hobbs}, {Lorimer}, {Lyne}  \&
  {Kramer}}{{Hobbs} et~al.}{2005}]{Hobbs:2005}
{Hobbs} G.,  {Lorimer} D.~R.,  {Lyne} A.~G.,   {Kramer} M.,  2005, \mn@doi
  [\mnras] {10.1111/j.1365-2966.2005.09087.x}, \href
  {http://adsabs.harvard.edu/abs/2005MNRAS.360..974H} {360, 974}

\bibitem[\protect\citeauthoryear{{Irrgang}, {Wilcox}, {Tucker}  \&
  {Schiefelbein}}{{Irrgang} et~al.}{2013}]{Irrgang:2013}
{Irrgang} A.,  {Wilcox} B.,  {Tucker} E.,   {Schiefelbein} L.,  2013, \mn@doi
  [\aap] {10.1051/0004-6361/201220540}, \href
  {http://adsabs.harvard.edu/abs/2013A%26A...549A.137I} {549, A137}

\bibitem[\protect\citeauthoryear{{Janka}}{{Janka}}{2013}]{Janka:2013}
{Janka} H.-T.,  2013, \mn@doi [\mnras] {10.1093/mnras/stt1106}, \href
  {http://adsabs.harvard.edu/abs/2013MNRAS.434.1355J} {434, 1355}

\bibitem[\protect\citeauthoryear{{Martin}, {Casares}, {Charles}, {van der
  Hooft}  \& {van Paradijs}}{{Martin} et~al.}{1995}]{Martin:1995}
{Martin} A.~C.,  {Casares} J.,  {Charles} P.~A.,  {van der Hooft} F.,   {van
  Paradijs} J.,  1995, \mnras, \href
  {http://adsabs.harvard.edu/abs/1995MNRAS.274L..46M} {274, L46}

\bibitem[\protect\citeauthoryear{{Miller-Jones}}{{Miller-Jones}}{2014}]{MillerJones:2014}
{Miller-Jones} J.~C.~A.,  2014, \mn@doi [\pasa] {10.1017/pasa.2014.7}, \href
  {http://adsabs.harvard.edu/abs/2014PASA...31...16M} {31, 16}

\bibitem[\protect\citeauthoryear{{Nelemans}, {Tauris}  \& {van den
  Heuvel}}{{Nelemans} et~al.}{1999}]{Nelemans:1999}
{Nelemans} G.,  {Tauris} T.~M.,   {van den Heuvel} E.~P.~J.,  1999, \aap, \href
  {http://adsabs.harvard.edu/abs/1999A%26A...352L..87N} {352, L87}

\bibitem[\protect\citeauthoryear{{Paczynski}}{{Paczynski}}{1990}]{Paczynski:1990}
{Paczynski} B.,  1990, \mn@doi [\apj] {10.1086/168257}, \href
  {http://adsabs.harvard.edu/abs/1990ApJ...348..485P} {348, 485}

\bibitem[\protect\citeauthoryear{{Remillard}, {Orosz}, {McClintock}  \&
  {Bailyn}}{{Remillard} et~al.}{1996}]{Remillard:1996}
{Remillard} R.~A.,  {Orosz} J.~A.,  {McClintock} J.~E.,   {Bailyn} C.~D.,
  1996, \mn@doi [\apj] {10.1086/176885}, \href
  {http://adsabs.harvard.edu/abs/1996ApJ...459..226R} {459, 226}

\bibitem[\protect\citeauthoryear{Repetto}{Repetto}{2015}]{Repetto:2015}
Repetto S.,  2015, private communication

\bibitem[\protect\citeauthoryear{{Repetto} \& {Nelemans}}{{Repetto} \&
  {Nelemans}}{2015}]{RepettoNelemans:2015}
{Repetto} S.,  {Nelemans} G.,  2015, preprint, \href
  {http://adsabs.harvard.edu/abs/2015arXiv150708105R} {} (\mn@eprint {arXiv}
  {1507.08105})

\bibitem[\protect\citeauthoryear{{Repetto}, {Davies}  \&
  {Sigurdsson}}{{Repetto} et~al.}{2012}]{Repetto:2012}
{Repetto} S.,  {Davies} M.~B.,   {Sigurdsson} S.,  2012, \mn@doi [\mnras]
  {10.1111/j.1365-2966.2012.21549.x}, \href
  {http://adsabs.harvard.edu/abs/2012MNRAS.425.2799R} {425, 2799}

\bibitem[\protect\citeauthoryear{{Urquhart}, {Figura}, {Moore}, {Hoare},
  {Lumsden}, {Mottram}, {Thompson}  \& {Oudmaijer}}{{Urquhart}
  et~al.}{2014}]{Urquhart:2014}
{Urquhart} J.~S.,  {Figura} C.~C.,  {Moore} T.~J.~T.,  {Hoare} M.~G.,
  {Lumsden} S.~L.,  {Mottram} J.~C.,  {Thompson} M.~A.,   {Oudmaijer} R.~D.,
  2014, \mn@doi [\mnras] {10.1093/mnras/stt2006}, \href
  {http://adsabs.harvard.edu/abs/2014MNRAS.437.1791U} {437, 1791}

\bibitem[\protect\citeauthoryear{{Willems}, {Henninger}, {Levin}, {Ivanova},
  {Kalogera}, {McGhee}, {Timmes}  \& {Fryer}}{{Willems}
  et~al.}{2005}]{Willems:2005}
{Willems} B.,  {Henninger} M.,  {Levin} T.,  {Ivanova} N.,  {Kalogera} V.,
  {McGhee} K.,  {Timmes} F.~X.,   {Fryer} C.~L.,  2005, \mn@doi [\apj]
  {10.1086/429557}, \href {http://adsabs.harvard.edu/abs/2005ApJ...625..324W}
  {625, 324}

\bibitem[\protect\citeauthoryear{{Wong}, {Valsecchi}, {Fragos}  \&
  {Kalogera}}{{Wong} et~al.}{2012}]{Wong:2012}
{Wong} T.-W.,  {Valsecchi} F.,  {Fragos} T.,   {Kalogera} V.,  2012, \mn@doi
  [\apj] {10.1088/0004-637X/747/2/111}, \href
  {http://adsabs.harvard.edu/abs/2012ApJ...747..111W} {747, 111}

\bibitem[\protect\citeauthoryear{{Wong}, {Valsecchi}, {Ansari}, {Fragos},
  {Glebbeek}, {Kalogera}  \& {McClintock}}{{Wong} et~al.}{2014}]{Wong:2014}
{Wong} T.-W.,  {Valsecchi} F.,  {Ansari} A.,  {Fragos} T.,  {Glebbeek} E.,
  {Kalogera} V.,   {McClintock} J.,  2014, \mn@doi [\apj]
  {10.1088/0004-637X/790/2/119}, \href
  {http://adsabs.harvard.edu/abs/2014ApJ...790..119W} {790, 119}

\makeatother
\end{thebibliography}

\label{lastpage}
\end{document}